\documentclass[11pt]{article}
\usepackage[a4paper,margin=1in]{geometry}
\usepackage{graphicx}
\usepackage{amsmath,amssymb}
\usepackage{booktabs}
\usepackage{hyperref}
\hypersetup{hidelinks}
\usepackage[numbers,sort&compress]{natbib}
\title{High-Resolution Reference Image Assisted Volumetric Super-Resolution of Cardiac Diffusion Weighted Imaging}
\author{Yinzhe Wu$^{*,a,b,c}$, Jiahao Huang$^{a,b,c}$, Fanwen Wang$^{a,b,c}$,\\ Pedro Ferreira$^{b,c}$, Andrew Scott$^{b,c}$, Sonia Nielles-Vallespin$^{b,c}$, Guang Yang$^{a,b,c}$\\[0.5em]
\small $^a$ Department of Bioengineering, Imperial College London, London SW7 2AZ, UK\\
\small $^b$ Cardiovascular Magnetic Resonance Unit, Royal Brompton Hospital, London SW3 6NP, UK\\
\small $^c$ National Heart and Lung Institute, Imperial College London, London SW7 2AZ, UK}
\date{}

\begin{document}
\maketitle

\begin{abstract}
Diffusion Tensor Cardiac Magnetic Resonance (DT-CMR) is the only in vivo method to non-invasively examine the microstructure of the human heart. Current research in DT-CMR aims to improve the understanding of how the cardiac microstructure relates to the macroscopic function of the healthy heart as well as how microstructural dysfunction contributes to disease. To get the final DT-CMR metrics, we need to acquire diffusion weighted images of at least 6 directions. However, due to DWI's low signal-to-noise ratio, the standard voxel size is quite big on the scale for microstructures. In this study, we explored the potential of deep-learning-based methods in improving the image quality volumetrically ($\times$4 in all dimensions). This study proposed a novel framework to enable volumetric super-resolution, with an additional model input of high-resolution b0 DWI. We demonstrated that the additional input could offer higher super-resolved image quality. Going beyond, the model is also able to super-resolve DWIs of unseen b-values, proving the model framework's generalizability for cardiac DWI super-resolution. In conclusion, we would then recommend giving the model a high-resolution reference image as an additional input to the low-resolution image for training and inference to guide all super-resolution frameworks for parametric imaging where a reference image is available.
\end{abstract}

\noindent\textbf{Index Terms---} Deep Learning, Super-resolution, Diffusion Weighted Imaging

\section{Introduction}
Cardiac Magnetic Resonance (CMR) imaging offers a non-invasive way to obtain multi-parametric myocardial tissue characterization. Diffusion Tensor CMR (DT-CMR), allows further examination of myocardial microstructure and re-orientation of cardiomyocytes within the heart during cardiac contraction \cite{Khalique2020}. To derive these diffusion tensors, we acquire Diffusion Weighted Imaging (DWI) of at least 6 directions and regressed them in reference to the B0-reference DWI to fit the tensors, which would then give a range of parametric maps to assist clinical diagnosis such as Mean Diffusivity (MD), Helix Angle (HA) and Fractional Anisotropy (FA).

However, the natural motion of the heart during the scan brings challenges to the image quality of DWIs acquired, causing its low Signal-to-Noise Ratio (SNR). As a result, both the low SNR and single-shot acquisitions lead to larger voxel sizes (usually 2.7$\times$2.7$\times$8 mm$^3$, length$\times$width$\times$slice thickness) for mitigation of low SNR in the DWI, where its slice thickness in particular is too big when considered over the scale for microstructures and compared to the usual thickness of the left ventricle myocardium. This also adversely enhances the partial volume and residual motion effects \cite{Gahm2014}, which could be harmful for clinical observations.

Recent works have demonstrated success of deep learning in spatial denoising \cite{Tian2022} and spatial super-resolution \cite{Kebiri2022} for MRI. This is usually done through a range of well-known deep learning network structures: ResNet, U-Net and generative adversarial network (GAN) \cite{Ledig2016}. However, these studies for MRI super-resolution focus either the in-plane resolution \cite{Tian2022} or the through-plane resolution \cite{Kebiri2022}, where no deep learning method has been validated for volumetric super-resolution (i.e., super-resolving both in-plane and through-plane resolution).

This study focuses on improving the spatial resolution of cardiac diffusion weighted imaging volumetrically. We demonstrated that U-Net could improve the resolution of cardiac DWI both in-plane and through-plane simultaneously by a factor of 4. In addition, the nature of DT-CMR as being a parametric imaging enabled us to have multiple observations of the imaged organ, we hypothesized that by having just the reference image acquired a high resolution it would boost the whole parametric image population for better quality through the deep learning model. Thus, we added an additional model input of high-resolution b0 DWI to the model and demonstrated that such addition could further boost its performance by giving the model a high-resolution reference. Going beyond, the model is also able to super-resolve DWIs of unseen b-values, proving the model framework's generalizability. This study is the first one to demonstrate the efficacy of deep learning models for volumetric super-resolution for MRI.

\section{Method}
\subsection{Dataset Acquisition and Pre-processing}
Data has been acquired from 10 healthy swine heart samples. These samples were frozen, defrosted and then fixed by either standard histology fix for various time periods (from 2 weeks to 6 months) prior to the scan. The variable time length of fixing allows us to observe different noise levels over the scan images, which is helpful for deep learning model training. The heart was suspended in Fomblin, where Fomblin is a contrast-free agent for MRI.

These datasets were acquired using a Siemens Vida 3 T MRI scanner (Siemens AG, Erlangen, Germany) with a diffusion-weighted spin-echo acquisition mode single-shot Echo-Planar Imaging (EPI) sequence with reduced phase field-of-view and fat saturation, Repetition Time (TR) = 25 sec, Echo Time (TE) = 73 msec, no acceleration, Echo Train Length (ETL) = 48, at a spatial resolution of 1.5 $\times$ 1.5 $\times$ 1.5 mm$^3$, Field Of View (FOV) of 72 $\times$ 120 mm$^2$, 5 averages. Diffusion was encoded in 256 directions with diffusion-weightings of b = 500 and 1000 sec/mm$^2$ in a short-axis slice to cover the whole heart, for 55-65 slices depending on the short-axis length of the heart. Additionally, reference images were also acquired with minimal diffusion weighting, named here as ``bref'' or ``b0'' images. The acquisition duration per each repetition of the whole series of diffusion weighted scans is approximately 3.5 - 4 hours, depending on the size of the heart.

In this study, we only focused on the mid-slices of the left ventricle, as other compartments have been collapsed in the fixation processed. To obtain the low-resolution slices downsampled volumetrically, each slice is firstly averaged with the neighboring 3 slices axially, achieving 4$\times$ through-plane downsampling. Each slice then underwent bilinear undersampling for in-plane downsampling. The pixel values of each slice were normalized to be mostly within the range [0, 1] prior to model input.

We split the 10 cases acquired into the training/validation/testing cases by the ratio 5:2:3, consisting of 1664:864:1120 2D slices respectively.

\subsection{Model Architecture}
We employed the U-Net variant in line with recent state-of-the-art generative image processing publication \cite{Nichol2021} as the backbone, where it added multiple multi-head self-attention layers to enhance the image quality of the results produced (Figure 1).

\begin{figure}[htbp]
\centering
\includegraphics[width=0.95\linewidth]{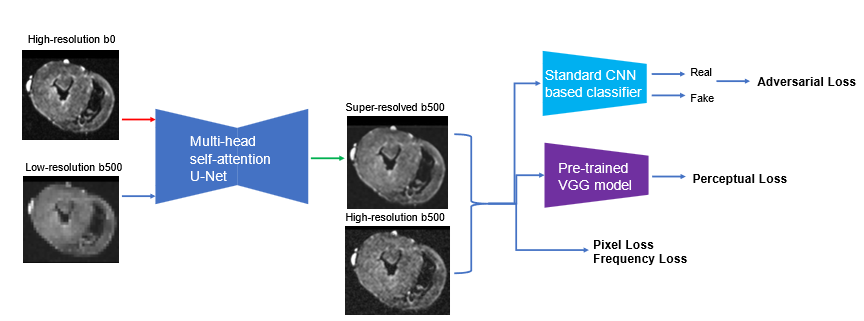}
\refstepcounter{figure}\par\small\noindent Figure~\thefigure. Proposed framework, where the model is supplied with an additional input to refer to the high-resolution (HR) b0 reference image while super-resolving the low-resolution image input.
\end{figure}

The main target of super-resolution was the b500 DWIs, where we input the model with the simulated 4$\times$ volumetrically downsampled ones as explained in 2.1. The main goal of investigation of this paper was to assess if an additional high-resolution reference given to the model as an additional input could improve its performance. To realize this, we concatenated the high-resolution b0 image scanned in reference to each b500 images with the low-resolution b500 image input channel-wise. The modified network took both images and then try to output the b500 image for super-resolution.

The model was designed to be trained in reference to the high-resolution DWI population through mean squared error function over both the image space and the frequency space (i.e., pixel loss and frequency loss) and a perceptual loss function, where the perceptual loss measured the latent space distance by a pre-trained VGG model. The model framework was also equipped with a standard CNN-based classifier as the discriminator to further enhance its super-resolution performance in an adversarial training strategy. The four loss functions (i.e., pixel loss, frequency loss, perceptual loss and adversarial loss) were added together linearly in a weighted way for model supervised learning.

\section{Experiment Setting}
\subsection{Implementation Details}
We trained the model with a batch size of 18 and a learning rate of 1e-4, halved every 20 epochs. We employed 3 NVIDIA GeForce 3090 GPUs for model training and inference. As discussed in 2.2, the input to the model were normalized 2-channel 2D slices, where they consisted of one channel of high-resolution b0 image and one channel of downsampled b500 image.

To establish the added benefit of the high-resolution b0 as an additional input for super-resolution, we conducted ablation study where compared the proposed framework against the conventional single channel input super-resolution framework. In addition, to showcase the generalizability of the model, we tested our proposed model over the images of an unseen b-value (1000) to see its results over this dataset.

We demonstrated the performance of model through 2 metrics: peak signal-to-noise ratio (PSNR) and structural similarity index metric (SSIM). These metrics were calculated per each 2D slice, with their means and standard deviations noted below.

\section{Result and Discussion}
The model was able to perform for the highest metrics in terms of PSNR and SSIM for the images of trained b-value (500) (Table 1 and Figure 2). The proposed method's result metrics were higher than the conventional approach (where no high-resolution reference image is supplied). In particular, in Figure 2 row 2, unlike the over-smoothing effects shown by the conventional method, the proposed method avoided such over-smoothing and showed more texture details in the result produced. This established the added benefit of supplying the additional high-resolution reference image for super-resolution tasks. We would thus recommend to include a high-resolution reference image as an additional input to the image to be super resolved/denoised for super-resolution tasks of parametric images where a reference image is available.

In addition, we observed that the model is able to super-resolve the images of unseen b-value (1000) (Table2 and Figure 3), further demonstrating the cross-b-value generalizability of the proposed framework.

\begin{table}[htbp]
\centering
\refstepcounter{table}\par\small\noindent Table~\thetable. Assess model's performance on trained b-value (500): Evaluation of the proposed framework (with both high-resolution (HR) b0 reference image and Low-Resolution (LR) input) against the conventional structure (where only LR input is supplied) through PSNR and SSIM metrics. Mean and standard deviation (std) of each metric population are supplied.\par\vspace{0.5em}
\small
\begin{tabular}{llll}
\toprule
Mean (std) & Bilinear & Proposed framework & Conventional framework \\
 & & (with HR b0 and LR) & (with LR only) \\
\midrule
PSNR ($\uparrow$) & 26.763 (0.824) & \textbf{32.615 (2.803)} & 29.151 (1.806) \\
SSIM ($\uparrow$) & 0.798 (0.061) & \textbf{0.863 (0.082)} & 0.816 (0.073) \\
\bottomrule
\end{tabular}
\end{table}

\begin{table}[htbp]
\centering
\refstepcounter{table}\par\small\noindent Table~\thetable. Assess model's performance on unseen b-value (1000): Evaluation of the proposed framework (with both high-resolution (HR) b0 reference image and low-resolution (LR) input) through PSNR and SSIM metrics. Mean and standard deviation (std) of the metric population are supplied.\par\vspace{0.5em}
\small
\begin{tabular}{lll}
\toprule
Mean (std) & Bilinear & Proposed framework \\
 & & (with HR b0 and LR) \\
\midrule
PSNR ($\uparrow$) & 26.726 (1.040) & \textbf{28.992 (2.365)} \\
SSIM ($\uparrow$) & 0.728 (0.063) & \textbf{0.762 (0.091)} \\
\bottomrule
\end{tabular}
\end{table}

\begin{figure}[htbp]
\centering
\includegraphics[width=0.96\linewidth]{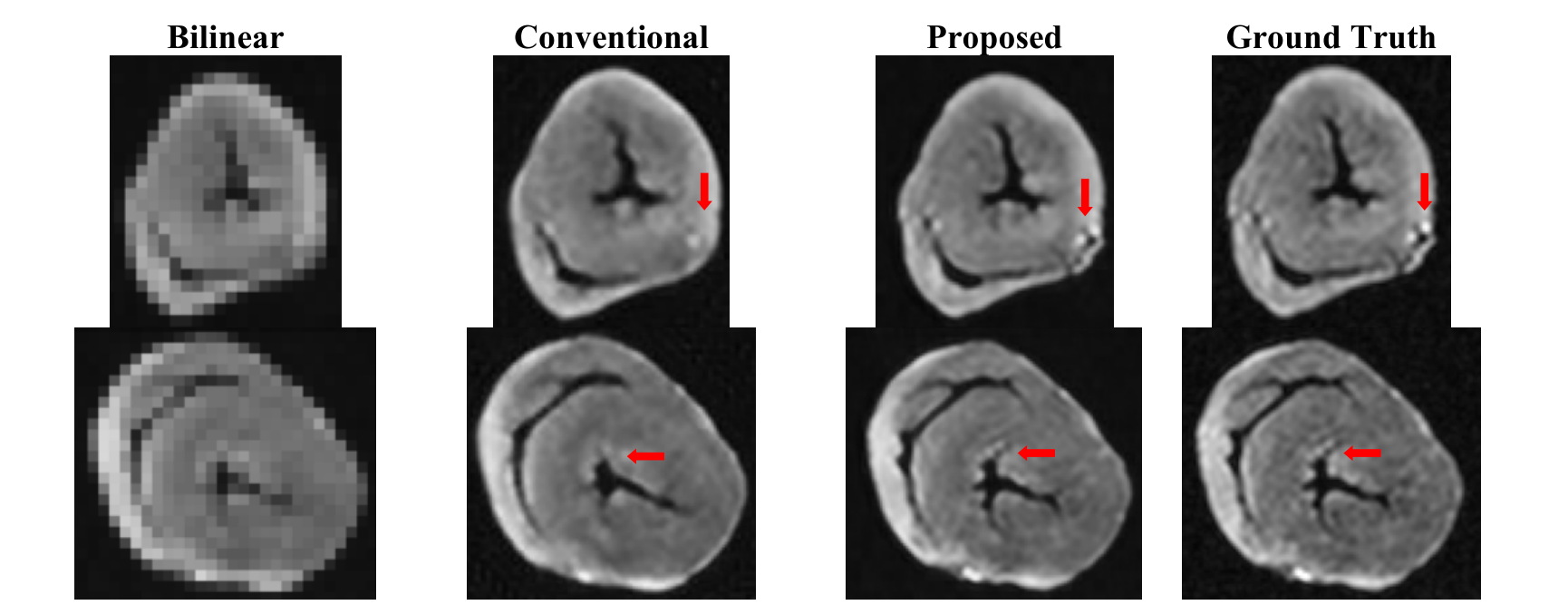}
\refstepcounter{figure}\par\small\noindent Figure~\thefigure. Assess model's performance on trained b-value (500): Sample b500 images of the bilinear interpolated, Ground Truth (GT) and the super-resolved images by the proposed and the conventional method.
\end{figure}

\begin{figure}[htbp]
\centering
\includegraphics[width=0.86\linewidth]{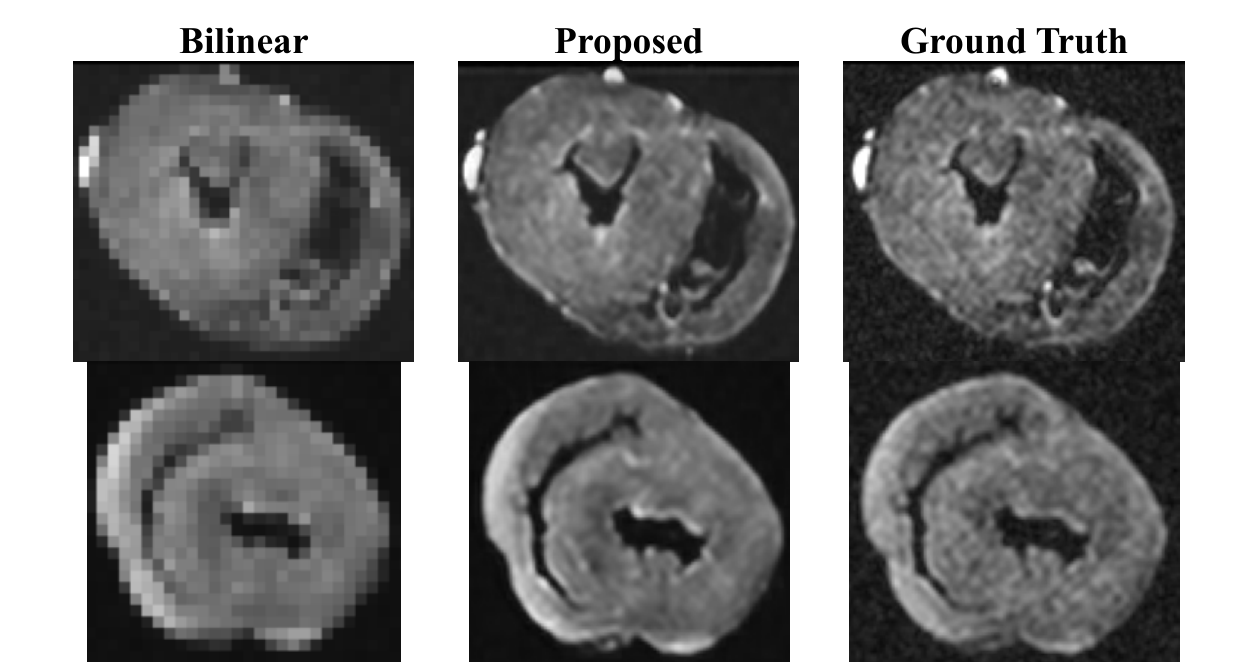}
\refstepcounter{figure}\par\small\noindent Figure~\thefigure. Assess model's performance on unseen b-value (500): Sample b500 images of the bilinear interpolated, Ground Truth (GT) and the super-resolved images by the proposed method.
\end{figure}

\section{Conclusion}
The super-resolution framework with an additional input of high-resolution b0 DWI can boost the performance of the network significantly in a generalizable way. We would then recommend giving the model the high-resolution reference image as a reference input in addition to low-resolution input for inference to guide all super-resolution frameworks for parametric imaging where a reference image is available.

\section{Funding Acknowledgment}
For the purpose of open access, the author(s) has applied a Creative Commons Attribution (CC BY) license to any Accepted Manuscript version arising.

This study is in part funded by Imperial College London President's PhD Scholarship and in part by UKRI Future Leaders
Fellowship (MR/V023799/1).

\bibliographystyle{unsrtnat}
\bibliography{references}

\end{document}